\documentclass[11pt]{article} 
\usepackage{epsfig,float}
\def\gsim{\;\raise0.3ex\hbox{$>$\kern-0.75em\raise-1.1ex\hbox{$\sim$}}\;}
\def\lsim{\;\raise0.3ex\hbox{$<$\kern-0.75em\raise-1.1ex\hbox{$\sim$}}\;}

\begin{document}

\begin{center}
{\large{\bf THE $\phi \rightarrow \gamma K^0 \bar{K^0}$ DECAY}}
\end{center}

\vspace{1cm}

\begin{center}
{\large{J.A. Oller}}
\end{center}

\begin{center}
{\small{\it Departamento de F\'{\i}sica Te\'orica and IFIC\\
Centro Mixto Universidad de Valencia-CSIC\\
46100 Burjassot (Valencia), Spain}}
\end{center}

\vspace{2cm}

\begin{abstract}
{\small{The branching ratio of the $\phi$ meson to $\gamma K^0 \bar{K^0}$ is 
calculated to be $5 \times 10^{-8}$ in a scheme which takes into account the 
different isospin channels involved, $I=0,1$ with the resonant, $f_0(980)$, 
$a_0(980)$, and non resonant contributions.}}  
\end{abstract}

\vspace{1.5cm}

The study of the process $\phi \rightarrow \gamma K^0 \bar{K^0}$ 
is an interesting subject since it provides a background to the reaction 
$\phi \rightarrow 
K^0 \bar{K^0}$. This latter process has been proposed as a way to study 
CP violating decays to measure the small ratio $\epsilon ' / \epsilon $ 
\cite{10C}, but since this implies seeking for very small effects a 
BR($ \phi \rightarrow \gamma K^0 \bar{K^0} $ )$ \gsim 10^{-6} $ will limit the 
scope of these perspectives. There are several calculations of this 
quantity \cite{1C,2C,3C,5C}. In \cite{4C} it is estimated for a {\it non 
resonant} decay process without including the $f_0$ and $a_0$ resonances. The 
issue is revisited in \cite{C}. 

Here, a different way to treat the scalar meson-meson 
sector, and its related $f_0 (980)$ and $a_0 (980)$ resonances, is proposed. 
For this we use a recent approach \cite{oller} to the S-wave meson-meson 
interaction for isospin $0$ and $1$ which reproduces the experimental data 
for those processes up to about $1.2$ $GeV$ and generates dynamically the 
$a_0$ and $f_0$ resonances. In this way, we will consider their interference 
and the energy 
dependence of their widths and coupling constants to the $K \bar{K}$ system. 
Furthermore, other possible contributions, non resonant, are also taken into 
account. The ideas and amplitudes exposed there were used 
in \cite{gamma} for the $\gamma \gamma \rightarrow \pi \pi$, $K \bar{K}$ and 
$\pi^0 \eta$ processes and a good agreement with the experiment was obtained. 

As in former works [1--5] we consider the process $\phi \rightarrow 
\gamma K^0 \bar{K^0}$ through an intermediate $K^+ K^-$ loop which 
couples strongly to the $\phi$ and the scalar resonances, see Fig.1.

\begin{figure}[h]

\centerline{
\protect
\hbox{
\psfig{file=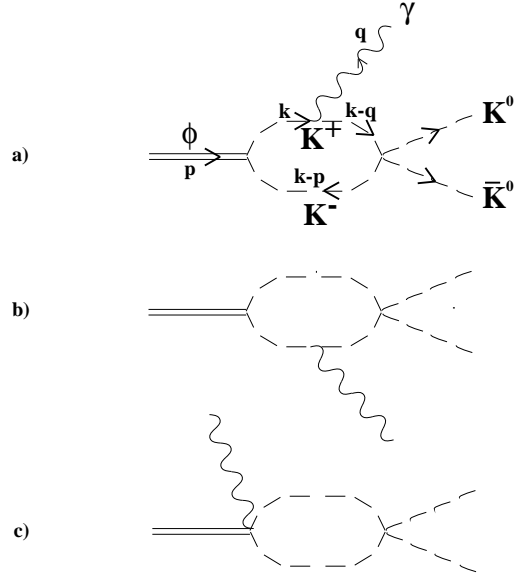,width=0.5\textwidth,angle=-90}}}
\caption{The loop radiation (a,b) and contact (c) contributions.}

\end{figure}

For calculating this loop contributions we use the minimal coupling to 
make the interaction between the $\phi$ and the $K^+ K^-$ 
mesons gauge invariant, then we have 

\begin{equation}
\label{Hint}
H_{int}=(e A_\mu + g_\phi \phi_\mu)i(K^+ \partial^\mu K^- - \partial ^{\mu}
K^+ K^-) - 2 e g_{\phi} A^{\mu} \phi_{\mu} K^+ K^-
\end{equation}      

Where $g_{\phi}$ is the coupling constant between the $\phi$ and the $K^+K^-$ 
system.

An essential ingredient to evaluate the loop in Fig.1 is the strong amplitude 
connecting $K^+ K^- $ with $K^0 \bar{K^0}$. As we said before we will use 
the amplitude calculated in \cite{oller}. This implies the sum of an infinite 
series of diagrams which is represented in Fig.2 for the diagram of 
Fig.1a, and the analogue corresponding to Figs.1b,c.

This series gives rise to the needed corrections due to final state 
interactions and in fact, from the vertex connecting the $K^+ K^-$ with 
the $K^0 \bar{K^0}$, this series is the same one that in \cite{oller} gives 
rise to the S-wave strong amplitude $K^+ K^- \rightarrow K^0 \bar{K^0}$. 
In this approach the vertex between 
the loops correspond to the lowest order chiral perturbation theory 
\cite{retaila}, $\chi PT$. Note that an analogous series before the loop 
with the emission of the photon is absorbed in the infinite series of 
diagrams contained in the $\phi$ resonance propagator. 

\begin{figure}[h]

\centerline{
\protect
\hbox{
\psfig{file=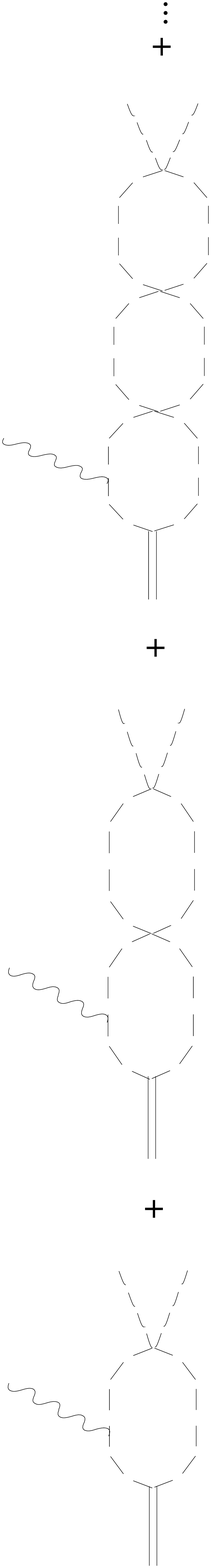,width=.13\textwidth,angle=-90}}}
\caption{Diagrammatic series which gives rise to the FSI from a general 
loop of Fig.1.}

\end{figure} 

First of all, let us see that the strong amplitude connecting $K^+ K^-$ with 
$K^0 \bar{K^0}$ calculated in the way shown in Fig.2 \cite{oller}  must 
factorize out of the integral.

For this consider the diagrams in Fig.1 but with the ${\cal O}(p^2)$ $\chi$PT 
amplitude connecting the kaons. This amplitude is given by

\begin{equation}
\label{potencial2a}
<K^0 \bar{K^0} |t| K^+ K^->= \frac{1}{2} [ t_{I=0} - t_{I=1} ]=-\frac{1}{4f^2}
[s+ \frac{4 m_K^2 - \sum_i p_i^2}{3}]
\end{equation}

where $f$ is the pion decay constant, $f\simeq 93$ $MeV$, $I$ refers to the 
isospin channel of the amplitude and the subindex $i$ runs from $1$ to 
$4$ and refers to any of the four kaons involved in the strong interaction. If 
the particle is on-shell then $p_i^2=m_K^2$. In our case $p_{K^0}^2=p_{\bar{
K^0}}^2=m_K^2$ so we have

\begin{equation}
\label{potencial2b}
-\frac{1}{4f^2}[s+\frac{(m_{K}^2-p_{K^+}^2)+(m_{K}^2-p_{K^-}^2)}{3}]
\end{equation}

The important point for the sequel is that the off-shell part, which 
should be kept inside the loop integration, will not contribute. 

In order to see this, note that, due to gauge invariance, 
the physical amplitude for $\phi \rightarrow \gamma K^0 \bar{K^0}$ has 
the form

\begin{equation}
\label{gaugea}
M(\phi (p) \rightarrow \gamma (q) K^0 \bar{K^0})=[g^{\mu \nu} (p \cdot q)-
p^\mu q^\nu]\epsilon_\mu ^\gamma \epsilon_\nu^\phi H(p\cdot q,Q^2,q\cdot Q)
\end{equation}

where $\epsilon_\mu ^\gamma$ and $\epsilon_\nu^\phi$ are the polarisation 
vectors of the photon and the $\phi$, $Q=p_{K^0}+p_{\bar{K^0}}$ and $H$ is 
an arbitrary scalar function. In 
the calculation of this loop contribution the problem is the presence of divergences 
in the loops represented in Fig.1. Following refs. \cite{1C,2C,3C} we will 
take into account the contribution of $p^\mu q^\nu$ of Figs. 1a,b, since 
Fig. 1c does not give such type of terms. Then, by gauge invariance, 
see formula (\ref{gaugea}), the coefficient for $(p\cdot q) g^{\mu \nu}$ is 
also fixed. In fact, as in ref. \cite{1C,2C,3C,C} it is shown, the $p^\mu q^\nu 
$ contribution will be finite since the off shell part of the strong amplitudes 
do not contribute, as we argue bellow, and then we are in the same situation 
than in the latter references. 

Take the diagrams of Figs.1a,b. These diagrams give the same contribution and this is the reason for the 
factor $2$ in front of the following integral accounting for both contributions.

\begin{equation}
\label{factora}
\begin{array}{l}
M'= \epsilon_\mu ^\gamma \epsilon_\nu^\phi \frac{2 e g_\phi }{i} 
\int \frac{d^4k}{(2 \pi)^4} \frac{(2k_\nu-p_\nu)(2k_\mu-q_\mu)}{
(k^2-m_K^2+i \epsilon )( (k-q)^2-m_K^2 + i\epsilon)( (k-p)^2-m_K^2+i \epsilon)}
  \\[2ex]
 \; \; \; \; \; \; \; \; \cdot \frac{-1}{4f^2}[Q^2+ \frac{(m_K^2-p_{K^+}^2)
 +(m_K^2-p_{K^-}^2)}{3}]
\end{array}
\end{equation} 

The momentum for each particle in the loop is indicated in Fig.1a and so 
we have that $p_{K^+}=k-q$, $p_{K^-}=k-p$. Concentrating in the 
off-shell part of the strong amplitude, we have the integral

\begin{equation}
\label{factorb}
\begin{array}{l}
\int \frac{d^4k}{(2\pi)^4} \frac{(2k_\nu-p_\nu)(2k_\mu-q_\mu)}{
(k^2-m_K^2+i\epsilon)((k-q)^2-m_K^2+i\epsilon)((k-p)^2-m_K^2+i\epsilon)} \cdot
[(k-q)^2-m_K^2 + (k-p)^2-m_K^2 ]=\\[2ex]
\int \frac{d^4k}{(2\pi)^4} \frac{(2k_\nu-p_\nu)(2k_\mu-q_\mu)}{
(k^2-m_K^2+i\epsilon)((k-p)^2-m_K^2+i\epsilon)} + \int \frac{d^4k}
{(2\pi)^4} \frac{(2k_\nu-p_\nu)(2k_\mu-q_\mu)}{(k^2-m_K^2+i\epsilon)
((k-q)^2-m_K^2+i\epsilon)} 
\end{array}
\end{equation}

Taking into account that 
\begin{equation}
\label{vector}
\epsilon_\mu^\phi \cdot p^\mu=0 \; ; \; \epsilon_\nu^\gamma \cdot q^\nu=0 \; 
\; (Feynman \; gauge)
\end{equation}

then we only have

\begin{equation}
\label{factorc}
\begin{array}{c}
\int \frac{d^4k}{(2\pi)^4}\frac{4 k_\mu k_\nu}{(k^2-m_K^2+i\epsilon)
((k-p)^2-m_K^2+i\epsilon)}+\int\frac{d^4k}{(2\pi)^4}\frac{4 k_\mu k_\nu }
{(k^2-m_K^2+i\epsilon)((k-q)^2-m_K^2+i\epsilon)}
\end{array}
\end{equation}

The above integrals do not give contribution to $q^\mu p^\nu$ since 
in each integral there is only one of the two vectors $q$ or $p$. In this 
way we see that the strong amplitude ${\cal O}(p^2)$ factorizes out on-shell in 
(\ref{gaugea}). Note that the important point in the former argumentation 
is the form of the off-shell part of the S-wave strong amplitude at 
${\cal O}(p^2)$ and this is common to any other S-wave meson-meson amplitude 
at this order, as one can see in \cite{oller}.  

Next we want to sum all the infinite series represented in Fig.2. The 
intermediate loops also contain $\pi \pi$ for $I=0$ and $\pi^0 \eta$ for 
$I=1$, since in ref.\cite{oller} coupled channel Lippmann-Schwinger equations 
were used with $\pi \pi$, $K \bar{K}$ in I=0 and $\pi^0 \eta$, $K \bar{K}$ 
in I=1. In ref. \cite{oller} it is shown that the meson-meson 
amplitude factorizes on-shell outside the loop integrals and since we have 
also here the ${\cal O}(p^2)$ strong amplitude factorizing we are then in the 
same situation as in \cite{oller} and we can substitute 
the ${\cal O}(p^2)$ strong amplitude by the one calculated to all orders 
in \cite{oller}. This result is an exact consequence of the 
approach used in \cite{oller}. Similar ideas were also used in 
\cite{gamma} to include the corrections coming from the final state 
interaction in $\gamma \gamma \rightarrow$meson-meson giving rise to a 
very good agreement with the experimental results. Then to all orders in 
the approach of \cite{oller} we have the amplitude

\begin{equation}
\label{strong}
t_S=\frac{1}{2}[t_{I=0}-t_{I=1}]
\end{equation}

Note that the amplitude obtained in \cite{oller} contains also the 
resonances $f_0(980)$ and $a_0(980)$ which are generated dynamically.

Then we have for the amplitude $ \phi(p) \rightarrow \gamma(q) K^0 \bar{K^0} $

\begin{equation}
\label{amplitude}
\begin{array}{c}
M = \epsilon_\mu^\gamma \epsilon_\nu^\phi \frac{2 e g_\phi }{i} 
t_{S}  \int \frac{d^4 k}{(2 \pi )^4} 
\frac{(2k_\nu -p_\nu )(2k_\mu -q_\mu )}
{(k^2-m_K^2+i\epsilon )((k-q)^2-m_K^2+i\epsilon )((k-p)^2-m_K^2+i\epsilon )}
\end{array}
\end{equation} 

This integral has been evaluated in \cite{1C} using dimensional regularization 
and confirmed in \cite{C}, with the result

\begin{equation}
\label{finalamplitude}
M=\frac{e g_\phi}{2 \pi^2 i m_K^2}\, I(a,b)\, [(p \cdot q)(\epsilon_\gamma 
\cdot \epsilon_\phi) - (p \cdot \epsilon_\gamma)(q\cdot \epsilon_\phi)]t_S
\end{equation}

with $a=M_\phi^2/m_K^2$ and $b=Q^2/m_K^2$

\begin{equation}
\label{Iab}
I(a,b)=\frac{1}{2(a-b)}-\frac{2}{(a-b)^2}(f(\frac{1}{b})-f(\frac{1}{a}))+
\frac{a}{(a-b)^2}(g(\frac{1}{b})-g(\frac{1}{a}))
\end{equation} 

where 

\begin{equation}
\label{fx}
\begin{array}{l}
f(x)=\left\{  
\vspace{-.5cm}
\begin{array}{cl}
-(arcsin(\frac{1}{2 \sqrt{x}}))^2  & x>\frac{1}{4}
\\ \frac{1}{4}[ln(\frac{\eta_+}{\eta_-})-i\pi]^2 & x<\frac{1}{4}
\end{array}
\right.
\\[4.5ex]
 g(x)=\left\{  
\vspace{-.5cm}
\begin{array}{c l}
(4x-1)^\frac{1}{2} arcsin(\frac{1}{2 \sqrt{x}}) & x>\frac{1}{4}
\\ \frac{1}{2}(1-4x)^\frac{1}{2}[ln(\frac{\eta_+}{\eta_-})-i\pi] & 
x<\frac{1}{4}
\end{array}
\right.
\\[4.5ex]
\eta_{\pm}=\frac{1}{2x}(1 \pm (1-4x)^{\frac{1}{2}})
\end{array}
\end{equation}

\vspace{0.5cm}

After summing over the final polarisations of the photon, averaging  
over the ones of the $\phi$ and taking into account the phase space for 
three particles \cite{PDG} one obtains

\begin{equation}
\label{widthform}
\Gamma (\phi \rightarrow \gamma K^0 \bar{K^0})=\int \frac{dm_{12}^2 dQ^2}
{(2 \pi)^3 192 M_{\phi}^3} |e g_{\phi} \frac{I(a,b)}{2 \pi^2 m_K^2}|^2
(M_\phi^2 - Q^2 )^2  |t_S|^2
\end{equation}

where $m_{12}^2=(q+p_{K^0})^2$. 

\vspace{0.3cm}

Taking $\frac{g_\phi^2}{4\pi}=1.66$ from its width to $K^+K^-$, $M_\phi=1019.41$
 $MeV$, $\Gamma (\phi)=4.43$ $MeV$, BR( $\phi \rightarrow K^0 \bar{K^0})=
0.34$ and using the mass of the $K^0$ for the phase 
space considerations, ref. \cite{PDG}, one gets 

\begin{equation}
\label{fullresult}
\begin{array}{rl}
\Gamma( \phi \rightarrow \gamma K^0 \bar{K^0})= & 2.22 \times 10^{-7} \, MeV
\\[2ex]
BR(\phi \rightarrow \gamma K^0 \bar{K^0})= & 0.50 \times 10^{-7}
\\[2ex]
\frac{ \Gamma( \phi \rightarrow \gamma K^0 \bar{K^0})}{\Gamma ( \phi 
\rightarrow K^0 \bar{K^0})}= & 1.47 \times 10^{-7}
\end{array}
\end{equation}

The uncertainties coming from the range of the possible values for the 
cut-off give a relative error around $20 \%$. 

Taking only into account the $I=0$ contribution

\begin{equation}
\label{noa0}
\begin{array}{rl}
\Gamma( \phi \rightarrow \gamma K^0 \bar{K^0})= & 8.43 \times 10^{-7} \, MeV
\\[2ex]
BR(\phi \rightarrow \gamma K^0 \bar{K^0})= & 1.90 \times 10^{-7}
\\[2ex]
\frac{\Gamma( \phi \rightarrow \gamma K^0 \bar{K^0})}
{\Gamma (\phi \rightarrow K^0 \bar{K^0})}= & 5.58 \times 10^{-7}
\end{array}
\end{equation}

and with only the $I=1$ 

\begin{equation}
\label{nof0}
\begin{array}{rl}
\Gamma( \phi \rightarrow \gamma K^0 \bar{K^0})= & 2.03 \times 10^{-7} \, MeV
\\[2ex]
BR(\phi \rightarrow \gamma K^0 \bar{K^0})= & 4.58 \times 10^{-8}
\\[2ex]
\frac{\Gamma( \phi \rightarrow \gamma K^0 \bar{K^0})}{\Gamma (\phi \rightarrow
 K^0 \bar{K^0})}= & 1.35 \times 10^{-7}
\end{array}
\end{equation}

We see that the process is dominated by the $I=0$ contribution and 
that the interference between both isospin channels is destructive.

From the former results we see that the $\phi \rightarrow \gamma K^0 \bar{K^0}$ 
background will {\bf not} be too significant for the purpose of 
testing CP violating decays from the $\phi \rightarrow K^0 \bar{K^0}$ process 
at DA$\Phi$NE in the lines of what was expected in \cite{C}. All these 
calculations have been done in a way that both the resonant and non-resonant 
contributions are considered at the same time and taking into account also 
the different isospin channels.

\vspace{.4cm}
 
{\bf Acknowledgements}

I would like to acknowledge useful discussions and a critical reading by 
E. Oset and M.J. Vicente-Vacas. I would also like to acknowledge financial 
support from the Generalitat Valenciana. This work is partially supported 
by CICYT, contract no. AEN 96-1719.


\vspace{.5cm}

\begin{thebibliography}{99}
\bibitem{10C} I.Dunietz, J.Hauser and J.L.Rosner, Phys. Rev. D35 (1987) 2166
\bibitem{1C} J.Lucio and J.Pestiau, Phys. Rev. D42 (1990) 3253; D43 (1991) 2447
\bibitem{2C} S.Nussinov and T.N.Truong, Phys. Rev. Lett. 63 (1989) 1349
\bibitem{3C} S.Nussinov and T.N.Truong, Phys. Rev. Lett. 63 (1989) 2003
\bibitem{5C} N.N.Achasov and V.N.Ivachenko, Nucl. Phys. B315 (1989) 465 
\bibitem{4C} N.Paver and Riazuddin, Phys. Lett. B246 (1990) 240
\bibitem{C} F.E.Close, N.Isgur, S.Kumano, Nucl. Phys. B389 (1993) 513
\bibitem{oller} J.A.Oller and E.Oset, Nucl. Phys. A620 (1997) 438
\bibitem{gamma} J.A.Oller and E.Oset , hep-ph/9706487, accepted for 
publication in Nucl. Phys. A
\bibitem{retaila} S.Weinberg, Phys. Rev. Lett. 19 (1967) 1264; J.Gasser and 
H.Leutwyler, Ann. of Phys. 158 (1984) 142
\bibitem{PDG}R.M.Barnett et al., Phys. Rev. D54 (1996)
\end{thebibliography}
\end{document}